\documentclass[aps,pre,longbibliography,notitlepage,unsortedaddress,floatfix,nofootinbib]{revtex4-1}

\usepackage[format=plain,labelfont={bf,small},textfont=small,justification=raggedright,singlelinecheck=false]{caption}
\usepackage{bm,amsmath,empheq,amssymb,graphicx,stmaryrd,float,subcaption,upgreek,sidecap,dsfont,mathtools,gensymb,multirow,array}
\usepackage[multiple]{footmisc}
\usepackage[abs]{overpic}
\usepackage{hyperref}

\usepackage{comment,footnote,mathrsfs,subcaption,xcolor}

\usepackage{tikz}
\usetikzlibrary{calc}

\newcommand{\bx}{{\bm x}}

\newcommand{\jump}[1]{{\left\llbracket #1 \right\rrbracket}}

\newcommand{\bc}{{\bm c}}

\newcommand{\bu}{{\bm u}}
\newcommand{\bl}{{\bm l}}
\newcommand{\bmo}{{\bm m}}

\newcommand{\bn}{{\bm n}}

\newcommand{\bE}{{\bf E}}

\newcommand{\bzero}{{\bf 0}}

\newcommand{\bp}{{\bm p}}
\newcommand{\bt}{{\bm t}}
\newcommand{\br}{{\bm r}}
\newcommand{\bd}{{\bm d}}

\newcommand{\G}{{G}}

\usepackage[utf8]{inputenc}
\usepackage[T1]{fontenc}
\usepackage{nomencl}
\graphicspath{{Figures/}}

\cleardoublepage           
\renewcommand{\nomname}{} 
\setcitestyle{numbers,square}

\begin{document}
\title{Adhesive tape loops}

\author{Krishnan Suryanarayanan}
\email{krishnan.s@iitgn.ac.in, krishnan@manit.ac.in}
\affiliation{Indian Institute of Technology Gandhinagar, Palaj, Gujarat-382055}
\author{Andrew B. Croll}
\email{andrew.croll@ndsu.edu}
\affiliation{North Dakota State University, Fargo, North Dakota-58102}
\author{Harmeet Singh}
\email{harmeet.singh@iitgn.ac.in}
\affiliation{Indian Institute of Technology Gandhinagar, Palaj, Gujarat-382055}
\date{\today}

\begin{abstract}
We present an experimental and theoretical study of the mechanics of an \emph{adhesive tape loop}, formed by bending a straight rectangular strip with adhesive properties, and prescribing an overlap between the two ends.
For a given combination of the adhesive strength and the extent of the overlap, the loop may unravel, it may stay in equilibrium, or open up quasi-statically to settle into an equilibrium with a smaller overlap.
We define the state space of an adhesive tape loop with two parameters: a non-dimensional adhesion strength, and the extent of overlap normalized by the total length of the loop.
We conduct experiments with adhesive tape loops fabricated out of sheets of polydimethylsiloxane (PDMS) and record their states.
We rationalize the experimental observations using a simple scaling argument, followed by a detailed theoretical model based on Kirchhoff rod theory.
The predictions made by the theoretical model, namely the shape of the loops and the states corresponding to equilibrium, show good agreement with the experimental data.
Our model may potentially be used to deduce the strength of self-adhesion in sticky soft materials by simply measuring the smallest overlap needed to maintain a tape loop in equilibrium.
\end{abstract}

\maketitle

\section{Introduction}

The mechanics of flexible solids joined by adhesion has long been a subject of interest to researchers, as evidenced by a large volume of work on the subject \cite{kendall1975,gent1977,bottega1991,johnson1971,boer1999,she2000,ghatak2005,glassmaker2004,vella2009,majidi2010,wilting2021}. 
Adhesives find utility in fields ranging from electronics to aerospace, and electrical appliances to a humble roll of adhesive tape.
Advent of technologies like MEMS and nanotubes \cite{glassmaker2004} 
along with interest in understanding and imitating adhesive mechanisms employed by biological creatures \cite{watson2015,stark2016}
have further motivated investigations into the mechanics of adhesion.

Therefore, estimating the force required to separate two solids attached by an adhesive is of immense practical interest, and critical for assessing which technologies are most competitive.  Sometimes the highest adhesion is desirable, for example in repairing acute damage to circulatory systems \cite{Yuk2019}, adhering automotive components to one another \cite{Fessel2007,Grujicic2008,Boutar2016}, or in construction \cite{vick1999,ciupack2017}.  On the other hand, there are many applications in which the lowest adhesion is desirable, such as on the hull of a ship which is easily be contaminated with organisms (below water) or ice (above water) \cite{lejars2012,Dhyani2021}.  There are even situations where strong adhesion is required, but only for a limited time.  Thus there has been some focus on measuring and imitating biological organisms which can easily accomplish this feat \cite{autumn2002,arzt2003,qu2008,kamperman2010,kim2008,ge2007}.    Regardless of the application, the measurement of the basic strength of an adhesive interaction requires an understanding of how applied forces make their way through an elastic body to the adhesive interface.

Large sections of the literature on the subject attempt to address the measurement question in the context of slender solids - such as elastic rods, plates, or membranes - attached to rigid substrates \cite{kendall1975,majidi2009,ponce2015,hanna2018,elder2020,Bartlett2023,Xie2025,Khatib2025}.
Adhesion between the slender solid and the rigid substrate in such works is treated as brittle (i.e. neglecting any cohesive zones), and incorporated in the model as an energy density required to advance a peeling front \cite{maugis1978,kinloch1994}.
Computing static configurations of such systems typically entails computing the shape of the non-adhered part of the solid, as well as the location and shape of the peeling front in the material for a given strength of the adhesive \cite{majidi2007}.
The adhered part of the solid takes up the shape of the substrate, which is known a priori.
Some early works on such problems are \cite{kendall1975,burridge1978}.

Computing equilibrium configurations of a slender solid which is under partial adhesive contact with itself is a problem more complicated than the aforementioned one, and also frequently turns up in applications \cite{twohig2024,glassmaker2004}.
The shape of the adhered region in such cases is unknown in general, and must be computed as part of the solution, in addition to the shape of the un-adhered region as well the location of the free boundary \cite{burridge1978}.
The existence of equilibrium is dependent upon not only the strength, but the extent of adhesion.
If either the strength or the extent falls short, maintaining equilibrium may become impossible.
Problems with self-adhesion in slender bodies considered in the literature are largely restricted to either symmetric configurations \cite{bottega1991,wilting2021}, 
or with a rigid substrate supporting the region of self-adhesion \cite{glassmaker2004}.
Both such arrangements render trivial the computation of the shape of self-adhered regions.

In this article, we consider a novel problem of self-adhesion without symmetry.
We consider a straight strip with finite thickness and a uniform rectangular cross-section, made from a material with adhesive properties, and bent into a loop with an overlap (Fig. \ref{fig:exp1}).
Henceforth, we will refer to such a loop as an \emph{Adhesive Tape Loop} (ATL).
Once the loop is allowed to relax for a reasonable duration of time, the following possibilities may arise: 1.) It may unravel and attain its natural straight configuration. 2.) It may stay as it is keeping the prescribed overlap length intact, (Fig.~\ref{fig:exp1} h) 3.) or, it may begin to open up and proceed quasi-statically to settle in a configuration with an overhang (Fig.~\ref{fig:exp1} e-g).
We study the mechanics of the ATL using experiments and theory.

We consider only planar static configurations of an ATL, and model the system using a planar version of Kirchhoff rod theory.
We base our analysis on balance laws of the Kirchhoff rod theory rather than variational principles widely used in the literature on adhesion problems  
\cite{majidi2010,wagner2013,napoligoriely2017,elder2020}.
We do not consider the possibility of delamination or blisters \cite{wagner2013,napoligoriely2017} 
developing along the adhered interface.
The opening of the loop is assumed to be initiated by the propagation of a peeling front nucleating at the end of the strip, in the event that the strength of the adhesion and the extent of the overlap are insufficient to establish equilibrium.
The overlapping region is subsequently modeled as an equivalent rod defined on the centerline of the inner region, with an effective bending modulus.
This converts an overlapping adhesive tape loop to an equivalent rod with a discontinuous bending modulus.

Our model reveals that upon appropriate normalizations the equilibrium of a tape loop, or the lack of it, is governed by three parameters: the thickness, the extent of overlap, and the strength of the adhesive.
For a given thickness of an ATL, the theory predicts a curve of limiting equilibria in the space of the overlap and adhesive strength, to be referred to as the state space, which separates the states that correspond to equilibrium from the ones that don't.
The theory also predicts the shape of the entire loop, including the adhered regions, and the internal forces and moments in the non-adhered region.
Furthermore, we track in the state space the paths of loops which open up from the imposed overlap and quasi-statically settle into equilibrium with a different overlap.
Predictions made from the theory show good agreement with experiments. 

The rest of the manuscript is organized as follows.
We begin by describing in section \ref{sec:experiments} the details of the experiments conducted, followed by a simple scaling model of the ATL in section \ref{sec:scalingmodel}.
The theoretical model of the ATL based on the Kirchhoff rod theory is detailed in \ref{sec:theoreticalmodel}. 
The complete boundary value problem governing the mechanics of the loop is stated in section \ref{sec:bvp}.
The results and their comparison with the experiments are discussed in section \ref{sec:results}, followed by conclusions in section \ref{sec:conclusions}.

\section{Experiments}\label{sec:experiments}

Sheets of polydimethylsiloxane (PDMS) were cast on pristine polycarbonate sheets obtained from one of several commercial sources.  Here we report primarily the results of Sylgard or Ecoflex PDMS formulations.  These materials come in two components which are mixed and then cured to form an elastomer.  The Sylgard system is slow to cure at room temperature, taking up to a week to fully crosslink.  Ecoflex, on the other hand, cures in approximately 10 minutes at room temperature.  Typically a solution would be mixed to a particular weight ratio (say 10 to 1 polymer to crosslinker with Sylgard), cast on the polycarbonate substrate and then rolled to a desired thickness with a Meyer bar.  Often, films of greater thickness were created through the addition of a new layer on top of one or several old layers.  A layer was added after the previous layer had cured to a tacky solid so that it would still crosslink with the newly added layer but was not pushed out by the Meyer bar.

\begin{figure}[t]
    \centering
    \includegraphics[width=0.95\linewidth]{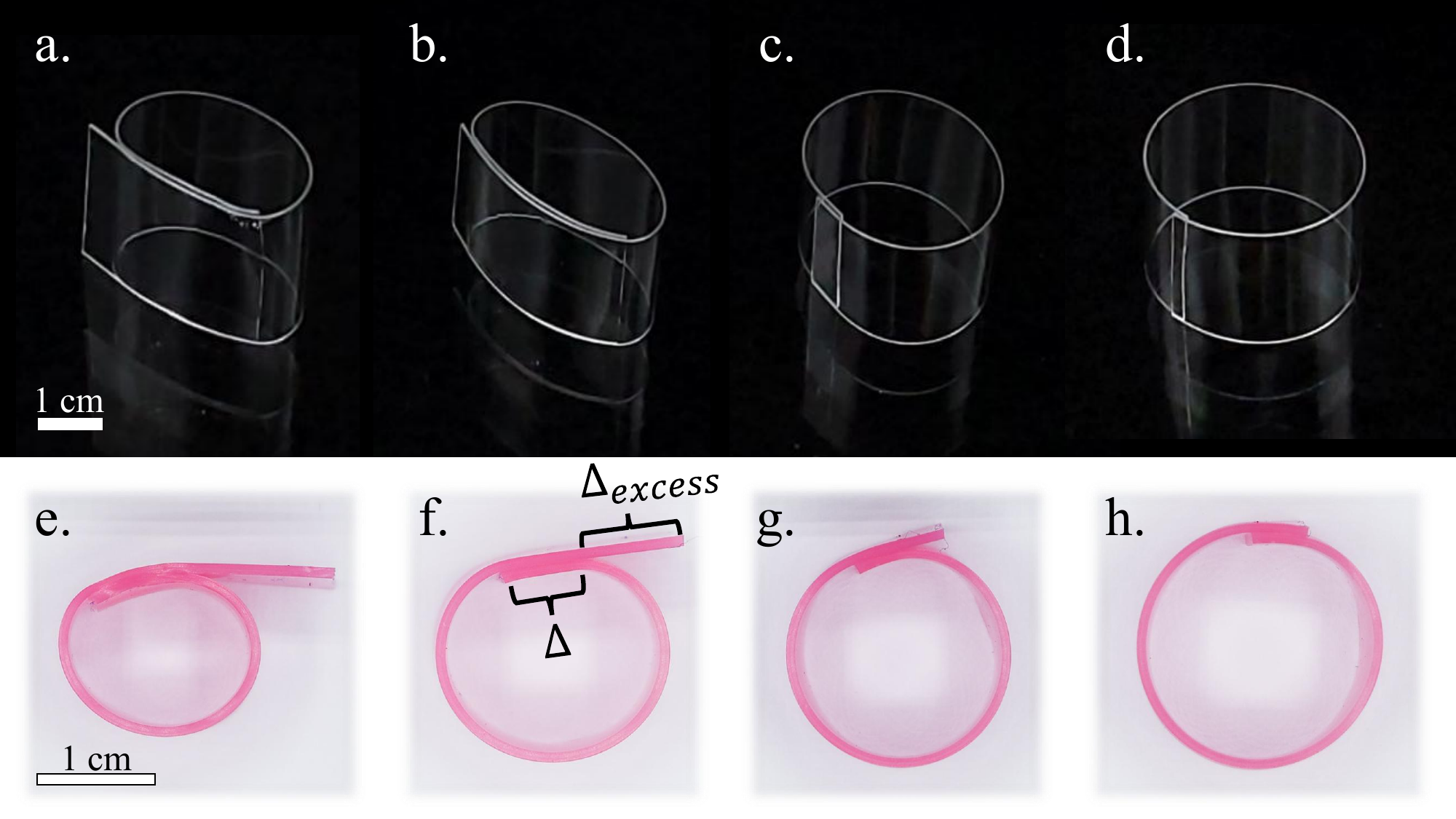}
    \caption{Typical experimental equilibrium loop shapes.  a.-d. A loop made from Sylgard 184 that is first in a marginal state, then (as the overlap is reduced) finds itself in a more circular shape.  e.-h. A loop made from Ecoflex which begins in a marginal state and progresses to a nearly circular shape as the overlap is reduced. }
    \label{fig:exp1}
\end{figure}

Once cured, sheets of elastomer are scored with a scalpel and ruler to create long strips of a desired width (typically 1~cm).  The width and length of the strip were then measured (in the flat state).  
Elastomer strips would then be manipulated into a loop shape through the use of tweezers or gloved hands.  Loops are placed on their side on a clean, rough surface for imaging (Fig.~\ref{fig:exp1}).  
Loops are tapped and lifted slightly with tweezers to ensure friction or adhesion with the substrate was not limiting the loop's ability to reach equilibrium.  Before measurement, loops were allowed to relax approximately 15 minutes.  While this does not ensure equilibrium is always perfectly attained, it strikes a reasonable balance between true equilibrium and the time needed to perform the number of measurements necessary.

Once equilibrated, a loop was imaged and image analysis was performed to extract a major and minor axis, a length of overlap ($\Delta>0$), and a length of unpeeled or excess loop denoted by $\Delta_{excess}>0$ (see Fig.~\ref{fig:exp1}).  Critically, the length of the entire loop between its start and the point of contact where the excess region begins is recorded.  Additionally, film thickness was assessed optically as well as with calipers (we favour the optical measurement due to the slight compression incurred by the calipers).

Typically, a single strip was used to form several loops.  Once a range of overlap lengths were explored, a loop would be shortened by cutting a small length from one end.  After shortening, loops were then created and imaged once again.  This process was repeated until such point as a loop could not be formed that would remain closed.  The end cut was alternated over the course of the experiment in order to minimize the overuse of any particular part of the strip.

To measure the self-adhesion of the loop, the racquet technique was used.  Here, a loop is folded back upon itself to form a tennis-racquet like shape.  The racquet is allowed to equilibrate and its width and length are measured optically.  Many excellent mechanical studies of the loop shape have been made \cite{py2007}, however, here we follow the work of Glassmaker and Hui \cite{glassmaker2004}.
In this case, the width of the loop can be written as $\mathcal{W}=1.25 \ell_{ea}$, where $\ell_{ea} = \sqrt{K/G_{c}}$ is the elasto-adhesive length. $K=Ewt^{3}/12(1-\nu^2)$ is the bending modulus of the strip with $\nu$ the Poisson ratio, Young's modulus $E$, width $w$, and thickness $t$.  The energy release rate, $G_{c}$ (roughly the work of adhesion per-unit length), can therefore be extracted from the racquet width given external knowledge of $E,\nu,$ and $t$.  Such measurements were performed at every point the loop’s overall length was shortened in order to detect any changes of adhesion strength due to the loops surfaces becoming contaminated during the coarse of a set of experiments.  The values determined in this manner were similar but slightly lower than those reported in \cite{vanDonselaar2023}, likely due to the slower speed of the racquet experiment.

\section{Scaling Model}\label{sec:scalingmodel}
A simple scaling approach can be used to estimate the limiting equilibrium of the loop closure problem.  This approach can only partially solve the problem, mainly due to assumptions about the loop geometry, which we assume to be circular with radius $R$ (and zero thickness).  By doing so, we have ignored curvature discontinuities due to adhesion at the inner and outer boundaries of the overlapping region, which are common in adhesion problems \cite{majidi2007,hanna2018,majidi2010,elder2020}.
The shortcomings of this assumption ultimately motivates the need for a more sophisticated approach which we discuss below.

The argument proceeds by assuming a crack were to open some small amount $\delta \ell$ between the outermost piece of the overlapping region and the inner region.  If it were to do so, it would release a small amount of mechanical energy due to the release of bending in this segment.  The crack would also reduce the energy stored in the interface by an amount proportional to the opened area.  Bending energy scales as $\sim(K/2R^2)\delta \ell$ and interfacial energy scales as $\sim G_c \delta \ell$ (where $G_c$ is the critical energy release rate, which is similar to Young-Dupr{\'e} work of adhesion).  Limiting equilibrium will occur if the two energies balance, which leads to the conclusion that $R\sim \ell_{ea}/\sqrt{2}$ at the tip of an arrested crack.  Here we use the elasto-adhesive length ($\ell_{ea}=\sqrt{K/G_c}$) to simplify the result.  

If a loop has a radius bigger than $ \ell_{ea}/\sqrt{2}$ it will maintain equilibrium, but if its radius is smaller than $ \ell_{ea}/\sqrt{2}$ it will open.  In the limit of the $\Delta \rightarrow 0$, the radius is approximately $L/2\pi$, where $L$ is the total length of the strip.  In other words, a loop must have a length larger than $ \sqrt{2} \pi \ell_{ea}$ to remain closed at all. Given the overlap length, $\Delta = L-2\pi R$, one finds a relation between total length and delta, which can be written:
\begin{equation}\label{Eqn:scaling1}
\frac{L^2}{\ell_{ea}^2} = 2\pi^2 \bigg{(}\frac{1}{1-\Delta/L}\bigg{)}^2.
\end{equation}
The unusual variable arrangement will be shown to be useful below.
The ratio on the left can also be rearranged as $G_cL^2/K$, and interpreted as a non-dimensional measure of the strength of the adhesive.

Equation \eqref{Eqn:scaling1} provides a good first estimate of the adhesive strength needed to maintain an adhesive tape loop in limiting equilibrium.
Next we develop a comprehensive model for an adhesive tape loop using Kirchhoff rod theory.

\section{Theoretical model}\label{sec:theoreticalmodel}
\begin{figure}[htbp]\label{eq:schematics}
		\captionsetup[subfigure]{justification=centering}
            \begin{subfigure}{0.345\textwidth}
				\includegraphics[width=\textwidth]{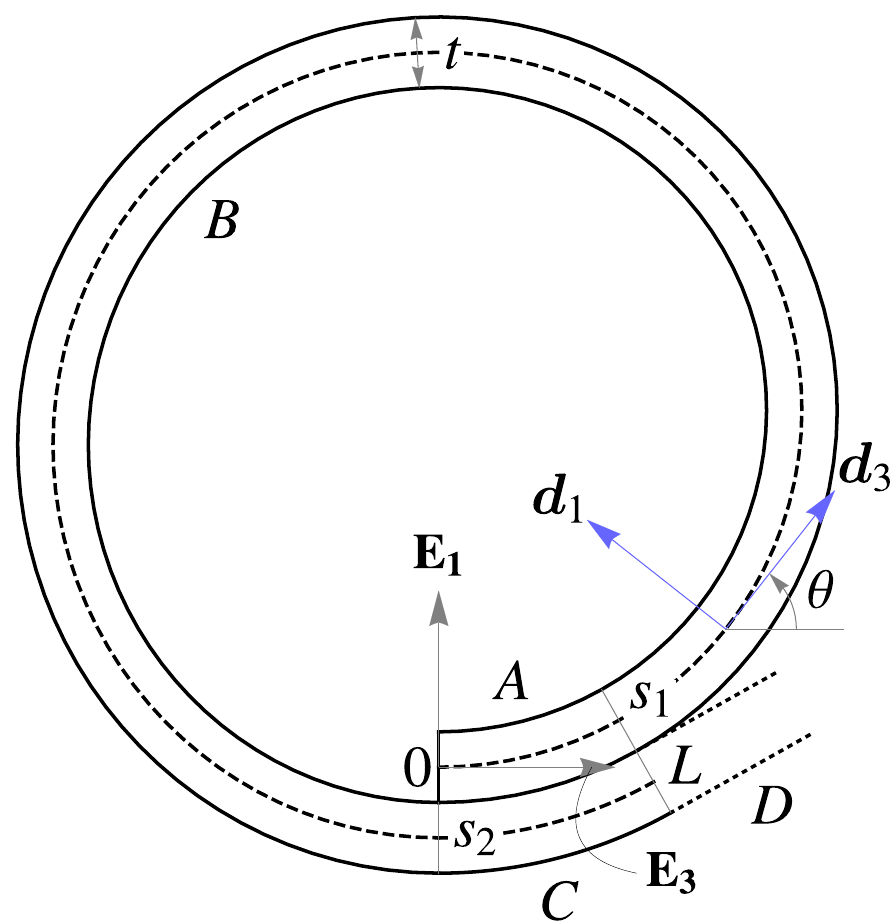}
				\caption{}
                \label{fig:schematic_a}
			\end{subfigure}
		\hspace{0.0cm}
	 	\begin{subfigure}{0.29\textwidth}
                \includegraphics[scale=0.3]{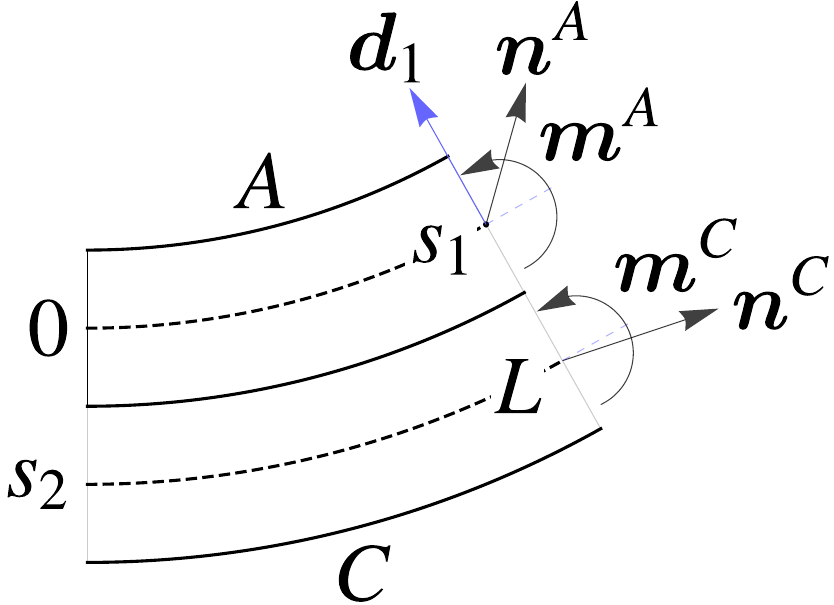}
				\caption{}
                \label{fig:schematic_b}
			\end{subfigure}
		\begin{subfigure}{0.345\textwidth}
				\includegraphics[width=0.95\textwidth]{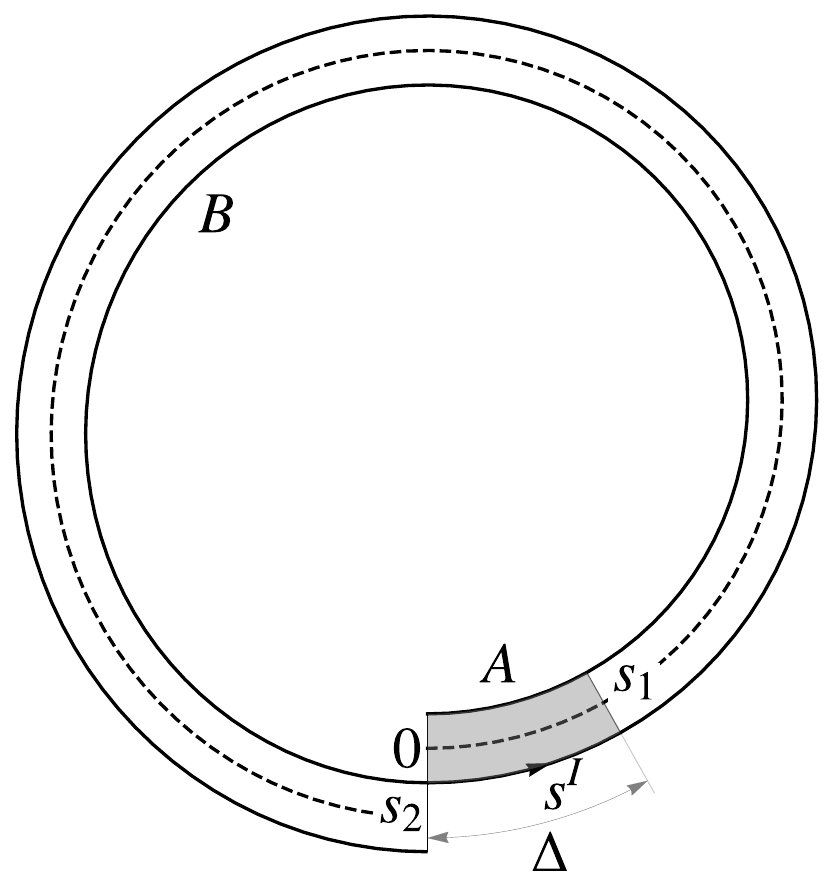}
				\caption{}
                \label{fig:schematic_c}
			\end{subfigure}
	\caption{(a) Schematic of a typical adhesive tape loop with overlap. (b) A free body diagram of the two overlapping regions $A$ and $C$. (c) A tape loop statically equivalent to the loop shown in a. with the overlapping regions $A$ and $C$ replaced by the shaded region with a modified constitutive relation.}
	\label{fig:schematics}
\end{figure}

Consider the schematics of a typical adhesive tape loop as shown in Fig.~\ref{fig:schematics}.
For the purpose of analysis, we divide the loop into four regions, namely regions A, B, C, and D.
We establish the governing equations valid individually in all these regions, and the appropriate jump/compatibility conditions that hold at their junctions.
Furthermore, we construct an equivalent system (Fig.~\ref{fig:schematic_c}) where regions $A$ and $C$ are to be treated effectively as a single rod with an modified constitutive law defined on the centerline of $A$.

\subsection{Kirchhoff rod theory}
In the Kirchhoff rod theory, a configuration of an elastic rod is identified with a centerline curve $\br\equiv\br(s)\in \mathbb{R}^3$, and a right-handed orthonormal frame of directors $\bd_i\equiv\bd_i(s)$, $i\in\{1,2,3\}$ \cite{antman1995}.
Here $s$ is the arc-length parameter of the elastic rod in some reference configuration.
The kinematics of a Kirchhoff elastic rod are represented mathematically by the following two relations,
\begin{align}
\br'=\bd_3\, ,\qquad \bd'_i = \bu\times\bd_i\, ,\label{eq:kinematics}
\end{align}
where the prime denotes derivation w.r.t. $s$.
The first condition above ensures that the centerline curve of the rod is inextensible and unshearable, while the second relation is a consequence of the orthonormality of the director frame.
The vector $\bu\equiv\bu(s)$ denotes the Darboux vector associated with the director frame, whose components $u_i\equiv u_i(s) := \bu\cdot\bd_i$ in the director basis represent the bending strains, $u_1$ and $u_2$, and the twist $u_3$, about $\bd_1,\bd_2,\bd_3$ respectively.

The local force and moment balances of an elastic rod takes the following form
\begin{subequations}\label{eq:force_moment_balance}
\begin{align}
     \bn' + \bp &= \bzero\, ,\label{eq:force_balance}\\
    \bmo' + \br'\times\bn + \bl &= \bzero\, \label{eq:moment_balance}.
\end{align}
\end{subequations}
Here $\bn\equiv\bn(s)$ and $\bmo\equiv\bmo(s)$ are the net internal force and moment exerted by the material at $s^+$ on the material in $s^-$, where $s^\pm = \lim_{\epsilon\rightarrow 0}(s\pm\epsilon)$ with $\epsilon>0$, across a cross-section centered at the arc-length coordinate $s$.
Furthermore, $\bp$ and $\bl$ denote, respectively, the external force and moment densities (per unit length) applied on the rod by its environment.
In the context of the adhesive tape loop, regions $B$ and $D$ will have $\bp=\bzero$ and $\bl=\bzero$, whereas regions $A$ and $C$ will experience a non-zero $\bp$ and $\bl$ due to the presence of adhesion.

The Kirchhoff rod theory further specifies the following constitutive relations, 
\begin{align}
    m_1 = K_1 u_1\, ,\qquad m_2=K_2 u_2\, ,\qquad m_3 = K_3 u_3 \, ,\label{eq:constitutive_relations}
\end{align}
where $m_i:=\bmo\cdot\bd_i$ are the components of the internal moment in the director basis, and $K_i$, with $i\in\{1,2,3\}$, are various bending moduli dependent of the material of the rod and the geometry of the cross-section.
The constitutive relations can alternatively be described by positing an energy density $W\equiv W(u_i)$ associated with the rod.
Relations \eqref{eq:constitutive_relations} are then equivalent to the following,
\begin{align}
    m_i = \frac{\partial W}{\partial u_i}\, ,\qquad W = \frac{1}{2}\sum_{i=1}^3 K_i u_i^2\, .\label{eq:energy_function}
\end{align}
The two relations above enable us to construct a conservation law that will prove to be useful in the subsequent analysis.
Taking the dot product of \eqref{eq:force_balance} and \eqref{eq:moment_balance} with $\br'$ and $\bu$ respectively, and adding the resulting scalar equations we obtain $\bn'\cdot\br' + \bmo'\cdot\bu + (\br'\times\bn)\cdot\bu + \bp\cdot\br' + \bl\cdot\bu=0$.
Using the product rule of derivatives to remove the derivatives from $\bn$ and $\bmo$ and subsequently using \eqref{eq:energy_function}$_1$, we arrive at the following,
\begin{align}
    \left(\bn\cdot\br' + \bmo\cdot\bu - W\right)' + \bp\cdot\br' + \bl\cdot\bu = 0\, .\label{eq:Hamiltonian_temp}
\end{align}
If the force and moment densities are constrained such that $\bp\cdot\br' + \bl\cdot\bu = 0$, we get the following conservation law upon integrating \eqref{eq:Hamiltonian_temp},
\begin{align}
    H(s) = H(0)\, ,\qquad\text{where}\qquad H(s) = \bn\cdot\br' + \bmo\cdot\bu - W\, .\label{eq:Hamiltonian_conservation}
\end{align}
Here $H\equiv H(s)$ will be referred to as the \emph{Hamiltonian} function \cite{maddocks1994,OReilly2017,singh2022,neukirch2026}. 

We will restrict ourselves to planar equilibria of an adhesive tape loop, meaning that we do not twist the strip before forming the loop.
Choosing the plane of deformation as the plane spanned by $\{\bd_1,\bd_3\}$, we state the following representations, to be invoked later, of various relevant fields,
\begin{subequations}\label{eq:planar_representations}
\begin{alignat}{3}
    & \br = r_1\bE_1 + r_3\bE_3\, ,\qquad && \bd_1 = \cos\theta\bE_1 - \sin\theta\bE_3\, ,\qquad &&\bd_3 = \sin\theta\bE_1 + \cos\theta\bE_3\\
    & \bn = n_1\bd_1 + n_3\bd_3\, ,\qquad && \bmo =K u_2 \bE_2\, ,\qquad && H = n_3 + \frac{1}{2} K u_2^2
\end{alignat}
\end{subequations}
where $\{\bE_1,\bE_3\}$ are a fixed Cartesian basis, and $\cos\theta:=\bd_3\cdot\bE_3$.
Since there is only one modulus $K_2$ that appears in the planar problem, we drop the subscript and replace it by $K\equiv K_2$ from this point onward.

\subsection{Problem setup}
\subsubsection{Force and moment balances}
We divide the entire tape loop into 4 parts (Fig.\ref{fig:schematics}): 1.) Region $A$, where $s^A\in[0,s_1]$, 2.) Region $B$, where $s^B\in(s_1,s_2)$, 3.) Region C, where $s^C\in(s_2,L)$, and finally, 4.) Region $D$, where $s^D\in(L,\infty)$.
We use superscripts to identify the region with which a particular function is associated.
The locations $s_1$ and $s_2$ on the centerline are free boundaries \cite{burridge1978}
whose values need to be determined as part of the solution.
The force and moment balance in regions $A$ and $C$ are written as
\begin{subequations}\label{eq:force_moment_balance_A_C}
    \begin{align}
        \frac{d\bn^A}{ds^A} + \bp^A & = \bzero\, , & \frac{d\bmo^A}{ds^A} + \frac{d\br^A}{ds^A}\times\bn^A + \bl^A&=\bzero\, ,\\
        \frac{d\bn^C}{ds^C} + \bp^C & = \bzero\, ,& \frac{d\bmo^C}{ds^C} + \frac{d\br^C}{ds^C}\times\bn^C + \bl^C&=\bzero\, ,
    \end{align}
\end{subequations}
where the adhesive interaction between the surfaces of the two regions is accounted for by the force and moment densities $\{\bp^A, \bp^C\}$ and $\{\bl^A, \bl^C\}$.
The force and moment balances for regions $B$ and $D$ are written as
\begin{subequations}\label{eq:force_moment_balance_B_D}
    \begin{align}
        \frac{d\bn^B}{ds^B}& = \bzero\, , & \frac{d\bmo^B}{ds^B} + \frac{d\br^B}{ds^B}\times\bn^B&=\bzero\, ,\label{eq:force_moment_balance_region_B}\\
        \frac{d\bn^D}{ds^D}& = \bzero\, ,& \frac{d\bmo^D}{ds^D} + \frac{d\br^D}{ds^D}\times\bn^D&=\bzero\, ,\label{eq:force_moment_balance_D}
    \end{align}
\end{subequations}
where no force and moment densities appear since the two regions interact with the rest of the loop only through their boundaries.
All field quantities are assumed to be functions of their respective arc-length coordinates in the region, e.g. $\bn^A\equiv\bn^A(s^A)$, etc.

We note that \eqref{eq:force_moment_balance_D}, together with the fact that region $D$ has a free end,
implies that $\bn^D(s^D)=\bzero$ and $\bmo(s^D)=\bzero$.
As a result, region $D$ does not interfere with the mechanics of the rest of the loop.

\subsubsection{Kinematic compatibility}\label{sec:kinematic_compatibility}
The force and moment balances given by \eqref{eq:force_moment_balance_A_C} and \eqref{eq:force_moment_balance_B_D} must be solved in conjunction with appropriate kinematic jump conditions at the internal boundaries, identified by the arc-length coordinates $\{s_1,s_2\}$, that ensures the continuity of the centerline and its tangents across the boundaries. 
These conditions can be succinctly represented as,
\begin{align}\label{eq:continuity_position_slope}
    \jump{\br}_{\{s_1,s_2,L\}} = \bzero\, ,\qquad \jump{\bd_3}_{\{s_1,s_2,L\}} = \bzero\, ,
\end{align}
where $\jump{\mathcal{A}}_{s_1} = \mathcal{A}^+ - \mathcal{A}^-$ represents the jump in the field $\mathcal{A}$ across $s=s_1$. 

Continuous contact between regions $A$ and $C$ requires that their centrelines must be uniformly separated by a distance $t$.
This kinematic constraint can be written as,
\begin{align}\label{eq:contact_kinematics_1}
    \br^C(s^C) = \br^A(s^A) - t\,\bd_1^A(s^A)\, .
\end{align}
Differentiating the above equation w.r.t. $s^A$, and using \eqref{eq:kinematics}$_2$, we obtain,
\begin{align}\label{eq:contact_kinematics_2}
    \frac{d\br^C}{ds^C}\frac{ds^C}{ds^A} = \frac{d\br^A}{ds^A} + t u_2^A \bd_3^A\, .
\end{align}
Noting that $d\br^C/ds^C = \bd_3^C(s^C)$, $d\br^A/ds^A = \bd_3^A(s^A)$, and $\bd_3^C(s^C)=\bd_3^A(s^A)$, we have the following relation from above,
\begin{align}\label{eq:contact_kinematics_3}
    \frac{ds^C}{ds^A} = 1+t\,u_2^A\, .
\end{align}
This differential equation governs the mapping between the material points on the centerlines of region $A$ and $C$. 
For a tape loop of zero thickness, the two arc-length coordinates coincide up to an additive constant.

\subsubsection{Equivalent single rod representation of the overlap}
Let us consider the region of overlap, comprising regions $A$ and $C$, in isolation from the rest of the loop (Fig.~\ref{fig:schematic_b}).
This combined region may be represented effectively as a single rod defined on the centerline of region $A$.
We will label various fields associated with this effective rod with an overbar as $\bar{()}$.
The effective internal force $\bar\bn^A\equiv\bar\bn^A(s^A)$, and moment $\bar\bmo^A\equiv\bar\bmo^A(s^A)$, of the equivalent rod are given by,
\begin{subequations}\label{eq:equivalent_internal_forces_moments}
    \begin{align}
        \bar\bn^A &= \bn^A + \bn^C\, ,\\
        \bar\bmo^A &= \bmo^A + \bmo^C - t\bd^A_1\times\bn^C\, .\label{eq:equivalent_moment}
    \end{align}
\end{subequations}
The equivalent rod does not experience external loads across its length, therefore, the equivalent force and moment balances for the effective rod can be written as,
\begin{subequations}\label{eq:equivalent_force_moment_balance}
    \begin{align}
        \frac{d\bar\bn^A}{ds^A} &= \bzero\, ,\label{eq:equivalent_force_balance}\\
        \frac{d\bar\bmo^A}{ds^A} + \frac{d\br^A}{ds^A}\times \bar\bn^A &= \bzero\, .\label{eq:equivalent_moment_balance}
    \end{align}
\end{subequations}

Equations \eqref{eq:equivalent_force_moment_balance} describe a rod defined on the centerline of $A$ that is statically equivalent to the overlapping region comprising regions $A$ and $C$.
To complete the description of this equivalent rod, we need an effective constitutive relation that relates the internal moment $\bar\bmo^A$ to the curvature $u_2^A$ of the centreline $\br^A$.
To this end, we use the constitutive relations of regions $A$ and $C$, i.e. $\bmo^A = K u_2^A\bd_2^A$ and $\bmo^C = K u_2^C\bd_2^C$, in \eqref{eq:equivalent_moment} to obtain,
\begin{align}\label{eq:equivalent_constitutive_relation_1}
    \bar\bmo^A = \left(K u_2^A + K u_2^C + t n_3^C\right)\bd_2^A\, .
\end{align}

Since the two centerlines $\br^A$ and $\br^C$ form a pair of parallel curves, their respective curvatures must satisfy the relation $1/u_2^C - 1/u_2^A = t$, using which we eliminate $u_2^C$ in favor of $u_2^A$ in \eqref{eq:equivalent_constitutive_relation_1}.
Furthermore, we assume that the force and moment densities $\bp^C$ and $\bl^C$ experienced by the centerline of region $C$ due to adhesion are constrained by the relation $\bp^C\cdot(\br^C)' + \bl^C\cdot\bu^C=0$.
This assumptions leads to the conservation of the associated Hamiltonian function \eqref{eq:Hamiltonian_conservation} in the region, whose planar representation $n_3^C + \tfrac{1}{2}K\left(u_2^C\right)^2 = H^C$, where $H^C$ is a constant, allows us to eliminate $n_3^C$ in favor of $H^C$ and $u_2^A$ in \eqref{eq:equivalent_constitutive_relation_1}.
Equation \eqref{eq:equivalent_constitutive_relation_1} then becomes,
\begin{align}\label{eq:equivalent_moment_constitutive}
    \bar\bmo^A = \bar m_2^A\bd_2^A\, ,\qquad \text{where} \qquad \bar m_2^A =  \left[ K u_2^A + K \frac{u_2^A}{1+t u_2^A} - \frac{t K}{2}\left(\frac{u_2^A}{1+t u_2^A}\right)^2 + H^C t\right]\, .
\end{align}
With this relation, we have reduced the overlapping region to a single effective rod whose internal moment $\bar m_2^A$ is related to the curvature $u_2^A$ of its centerline by the above expression.
Relation \eqref{eq:equivalent_moment_constitutive} agrees with equation (104) of \cite{starostin2014} which states the equivalent constitutive relation for two planar elastic rods of finite thicknesses joined together.

Finally for the equivalent loop, the following jump conditions on the internal force and moment must hold,
\begin{align}\label{eq:force_moment_jump_s1}
      \jump{\bn}_{s_1}=\bzero\, ,\qquad\jump{\bmo}_{s_1}=\bzero\, ,
\end{align}
which denote the continuity of the internal force and moment across the point $s=s_1$ of the effective rod.


\subsubsection{Determination of the free boundaries}
A typical experiment done with ATLs prescribes the overlap $\Delta$ along the adhered surfaces of regions $A$ and $C$.
Prescribing $\Delta$ renders the location of the point $s_1$ and $s_2$ unknowns of the problems.
We refer to these points as \emph{free boundaries} \cite{burridge1978}.
These additional unknowns are computed as follows.

Let $s^I$ be the arc-length coordinate along the adhered interface of region $A$ and $C$.
The interface is thus a curve parallel to the centerline of region $A$.
Using arguments similar to the ones in section \ref{sec:kinematic_compatibility}, $s^I$ can be related to $s^A$ as follows,
\begin{align}
    \frac{ds^I}{ds^A} = 1+\frac{t}{2}u_2^A\, ,\qquad s^I(0) = 0\, ,\qquad s^I(s_1) = \Delta\, ,\label{eq:interfacial_arc_length}
\end{align}
where the two boundary conditions fix the location of the overlap in the loop.
Inclusion of \eqref{eq:interfacial_arc_length} in the full boundary value problem would enable the determination of $s_1$.

The location of the second free boundary, i.e. $s=s_2$ will be obtained by assuming the continuity of the Hamiltonian function across $s_2$,
\begin{align}
    \jump{H}_{s_2} = 0\, .\label{eq:hamiltonian_jump_s2}
\end{align}
We state this jump condition here without derivation.
This condition essentially ensures that the $s_2$ is located such that the energy of the system remains stationary w.r.t. kinematically admissible perturbations to $s_2$.
For problems that admit a variational structure, this condition is also known as the \emph{Weierstrass-Erdmann} corner condition \cite{gelfand2000}.
We do not dwell further on it.
Explanations and justifications, from a variational perspective, of this jump condition can be found in the literature \cite{OReilly2017,singh2022,hanna2018}.
A comprehensive survey on the topic can be found in \cite{majidi2007,neukirch2026}.

\section{The boundary value problem}\label{sec:bvp}
The theory developed in the previous sections can now be used to pose a boundary value problem that governs the equilibrium of the equivalent ATL (Fig.\ref{fig:schematic_c}).
We non-dimensionalise all lengths in our system by $L$, and forces by $K/L^2$.
To avoid clutter in notation, we refrain from using a different notation for the scaled entities.
The full non-dimensional boundary value problem thus obtained is stated as follows,
\begin{subequations}\label{eq:bvp}
\begin{align}
        &\frac{dr_1^A}{ds^A}=\sin\theta^A\, ,& &r_1^A(0)=0\, ,\label{eq:bvp_A_r1}\\
        &\frac{dr_3^A}{ds^A}=\cos\theta^A\, ,& &r_3^A(0)=0\, ,\label{eq:bvp_A_r3}\\
        &\frac{d\theta^A}{ds^A}=u_2^A\, ,& &\theta^A(0)=0\, ,\label{eq:bvp_A_theta}\\
        &\frac{d\bar{n}_1^A}{ds^A}=-u_2^A\bar{n}_3^A\, ,& &r_1^A(s_1)=r_1^B(s_1)\, ,\label{eq:bvp_A_n1}\\
        &\frac{d\bar{n}_3^A}{ds^A}=u_2^A\bar{n}_1^A\, ,& &r_3^A(s_1)=r_3^B(s_1)\, ,\label{eq:bvp_A_n3}\\
        &\frac{du_2^A}{ds^A}=\frac{-\bar{n}_1^A}{\left(1 + \frac{1}{1+t u_2} - \frac{2 t u_2}{(1+t u_2)^2} + \frac{t^2 u_2^2}{(1+t u_2)^3}\right)}\, ,& &\theta^A(s_1)=\theta^B(s_1)\, ,\label{eq:bvp_A_u2}\\
         &\frac{dr_1^B}{ds^B} = \sin\theta^B\, ,& &\bar{n}_1^A(s_1) = n_1^B(s_1)\, , \label{eq:bvp_B_r1}\\
         &\frac{dr_3^B}{ds^B} = \cos\theta^B\, ,& &\bar{n}_3^A(s_1) = n_3^B(s_1)\, ,\label{eq:bvp_B_r3}\\
         &\frac{d\theta^B}{ds^B} = u_2^B\, ,& &\bar m_2(s_1) = u_2^B(s_1)\, ,\label{eq:bvp_B_theta}\\
         &\frac{dn_1^B}{ds^B} = - u_2^B n_3^B\, , & &r_1^B(s_2) = -t\, ,\label{eq:bvp_B_n1}\\
         &\frac{dn_3^B}{ds^B} = u_2^B n_1^B\, , & &r_3^B(s_2) = 0\, ,\label{eq:bvp_B_n3}\\
         &\frac{du_2^B}{ds^B} = -n_1^B\, , & &\theta^B(s_2) = 2\pi\, ,\label{eq:bvp_B_u2}\\
         &\frac{ds^C}{ds^A} = 1 + t u_2^A\, ,& &s^C(s_2) = s_2\, ,\label{eq:bvp_sC}\\
         &\frac{ds^I}{ds^A}=1+ \frac{t}{2} u_2^A\, , & & s^C(1) = 1 \, , \label{eq:bvp_sI}\\
         &\frac{ds_1}{ds^A} = 0\, , & &s^I(0) = 0\, ,\label{eq:bvp_s1}\\
         &\frac{ds_2}{ds^A} = 0\, , & &s^I(s_1)=\Delta \, ,\label{eq:bvp_s2}\\
         &\frac{dH^C}{ds^A} = 0\, , & &H^C(s_2) = n_3^B(s_2) + \frac{1}{2}u_2^B(s_2)^2\, .\label{eq:bvp_HC}
\end{align}
\end{subequations}
Equations \eqref{eq:bvp_A_r1} and \eqref{eq:bvp_A_r3} are the $\bE_1$ and $\bE_3$ components of \eqref{eq:kinematics}$_1$, while \eqref{eq:bvp_A_u2} is the $\bd_2$ component of \eqref{eq:equivalent_moment_balance} substituted with \eqref{eq:equivalent_moment_constitutive}.
Equations \eqref{eq:bvp_A_n1} and \eqref{eq:bvp_A_n3}, are the $\bd^A_1$, $\bd^A_3$ components of the equivalent force balance \eqref{eq:equivalent_force_balance}, whereas \eqref{eq:bvp_A_u2} is the $\bd^A_2$ component of \eqref{eq:equivalent_moment_balance} substituted with \eqref{eq:equivalent_moment}. 
Equations \eqref{eq:bvp_B_r1} to \eqref{eq:bvp_B_u2} similarly correspond to the kinematics and force and moment balance equations governing region $B$.
Equations \eqref{eq:bvp_sC} and \eqref{eq:bvp_sI} are simply \eqref{eq:contact_kinematics_3} and \eqref{eq:interfacial_arc_length}, and \eqref{eq:bvp_s1} to \eqref{eq:bvp_HC} are trivial equations used to convert the unknown parameters $s_1,\, s_2,$ and $H^C$ as constant functions to be determined as part of the solution.

We note that the scaled BVP is independent of $K$, implying that the shape of the equivalent ATL, described by the functions $\{r^A_1,r^A_3, r^B_1, r^B_3\}$, is independent of the bending stiffness of the loop.

Solving the nonlinear boundary value problem \eqref{eq:bvp} entails prescribing values of $t$ and $\Delta$ and computing the unknown vector $\{r_1^A, r_3^A, \theta^A, \bar n_1^A, \bar n_3^A, u_2^A, r_1^B, r_3^B, \theta^B, n_1^B, n_3^B, u_2^B, s^C, s^I, s_1, s_2, H^C\}$.
We numerically solve the BVP for various values of $t$ and $\Delta$ using the \texttt{NDSolve} function of \texttt{Mathematica, Version 12.0.0.0}, which employs a shooting method in the background.
For the ease of executing a numerical solution, we map various arclengths to a single parameter $\xi\in[0,1]$ as follow: $s^A = s_1\xi\, ,s^B = (s_2-s_1)\xi + s_1\, ,$  and $ s^C = (1-s_2)\xi+s_2\, ,$ before feeding it to \texttt{NDSolve}.

A solution to the boundary value problem above delivers the equilibrium configuration of the equivalent rod depicted in Fig.~\ref{fig:schematic_c} for a given value of $\Delta$.
The centerline of region $C$ can then be constructed using \eqref{eq:contact_kinematics_1}, which completes the shape of the original loop.
Once the equilibrium is obtained, the critical adhesive strength $G_c$ required to maintain such an equilibrium is obtained using the standard adhesion boundary condition $G_c=\jump{(m_2^A)^2/2K}$, which states that at the limiting equilibrium, the jump in the bending energy density across $s_1$ must be exactly compensated by the strength of the adhesion \cite{majidi2007,hanna2018,majidi2010,elder2020}.

\begin{figure}[t]
\captionsetup[subfigure]{justification=centering}
	\begin{subfigure}{0.33\textwidth}
		\includegraphics[width=0.975\textwidth]{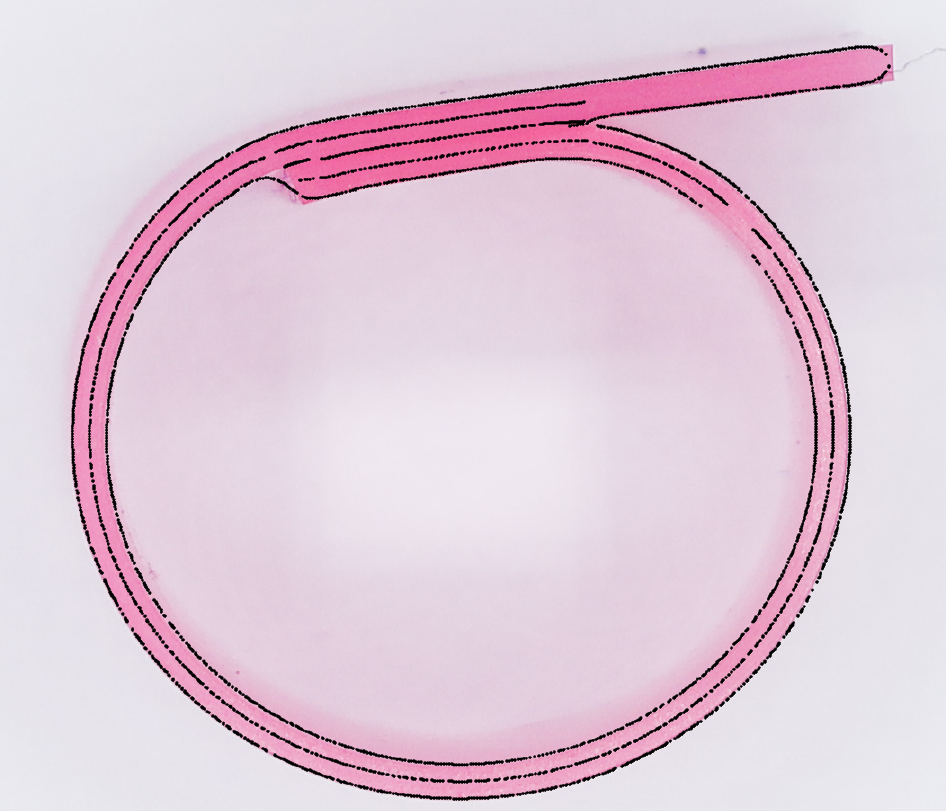}
		\caption{}
		\label{fig:shape_comparison_a}
	\end{subfigure}
	\begin{subfigure}{0.33\textwidth}
		\includegraphics[width=0.85\textwidth]{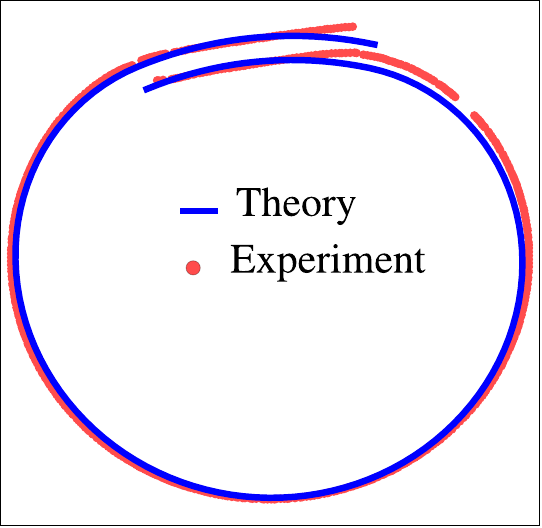}
		\caption{}
		\label{fig:shape_comparison_b}
	\end{subfigure}
	\begin{subfigure}{0.325\textwidth}
		\includegraphics[width=\textwidth]{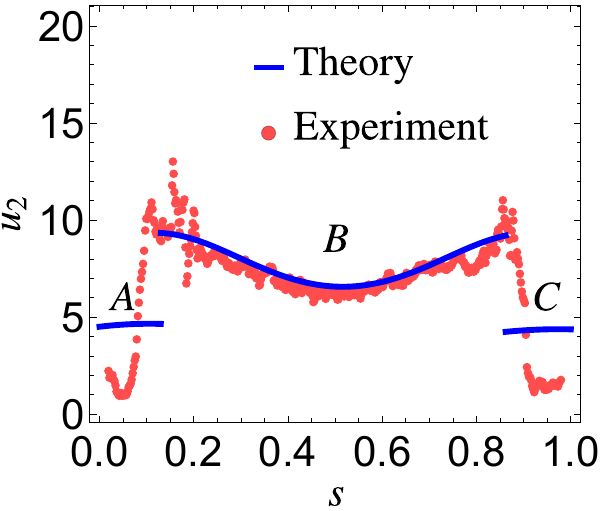}
		\caption{}
		\label{fig:shape_comparison_c}
	\end{subfigure}        
	\caption{(a) Overlay of the boundary of tape loop obtained by image processing and computed centerline (black) with the experiment,   (b) Comparison of the centerline prediction from theory with experiment and (c) Comparison of the curvature plots obtained from theory and experiment for a single sample.}
	\label{fig:shape_comparison}
\end{figure}

\section{Results}\label{sec:results}
Over the course of the experiments, strips of different lengths and thicknesses were bent to form a loop by imposing an overlap length denoted by $\Delta$.
We classify such imposed states by the non-dimensional overlap length and strength of adhesion, denoted by the pair $\{\Delta/L,\G_cL^2/K\}$.
For strips of varied thicknesses, imposing a state resulted in three outcomes in general. 

First, a loop could spontaneously open after being closed, returning to its unbent state.
In this case, we describe the imposed state as corresponding to \emph{no-equilibrium}.
Second, the imposed state could hold itself together as is, in which case we describe the state as corresponding to \emph{equilibrium}.
Finally, a loop could slowly open up from its initially imposed overlap length $\Delta_{initial}$, and settle to a final overlap of $\Delta$.
In this case, we refer to the final state as \emph{equilibrium with overhang}.
We denote the excess peeled of length as $\Delta_{excess}$, so that $\Delta_{initial} = \Delta + \Delta_{excess}$.

For a strip of given length $L$ and thickness $t$, the three states would follow the order described above as $\Delta$ was increased.  
At short overlaps, loops found no equilibrium.  
As the overlap increased the loops could maintain equilibrium in the states imposed on them.  
As the overlap was increased further, equilibrium with overhang would be reached.  The length of overlap where changes in state were observed was found to depend on the length and thickness of the strip. As a strip was made shorter, larger $\Delta$ would be needed to maintain equilibrium.  
As a strip were made thicker, the \emph{no-equilibrium} region persisted to larger overlaps.  
We also observe re-entrant behaviour in certain situations.  
For example, some thin strips would progress from \emph{no-equilibrium} to \emph{equilibrium} to \emph{equilibrium with overhang}, then find \emph{equilibrium} and finally \emph{equilibrium with overhang} as $\Delta$ was increased.

Before we rationalize these experimental observations using our theoretical model, we first validate the theory by comparing equilibrium shapes obtained from \eqref{eq:bvp} with the shapes of the experimental loops.
Fig.~\ref{fig:shape_comparison_a} shows an experimental loop with an overhang.
The inner and outer boundaries of the loop, extracted from the picture using a procedure detailed in Appendix~\ref{app:image_processing}, along with the computed centerline is overlayed on the picture.
Fig.~\ref{fig:shape_comparison_b} shows good agreement between the theoretical prediction of the centerline with the experiment.
Fig.~\ref{fig:shape_comparison_c} shows a comparison of the curvatures predicted by the theory with the curvatures computed from the experimental centerline. 
The agreement between theoretical and experimental curvatures appears good in region $B$ of the tape loop.
As predicted by the theory, the experimental curvatures display a sharp change in curvatures at the boundaries of region $B$, although the magnitudes of the change does not match well.
This discrepancy could be attributed to higher shear deformation in regions $A$ and $C$, as compared to region $B$, which makes the assumption of inextensibility and unshearability a poor approximation of the deformation in the overlapping region.

\begin{figure}[t]
        \centering
        \includegraphics[width=0.98\textwidth]{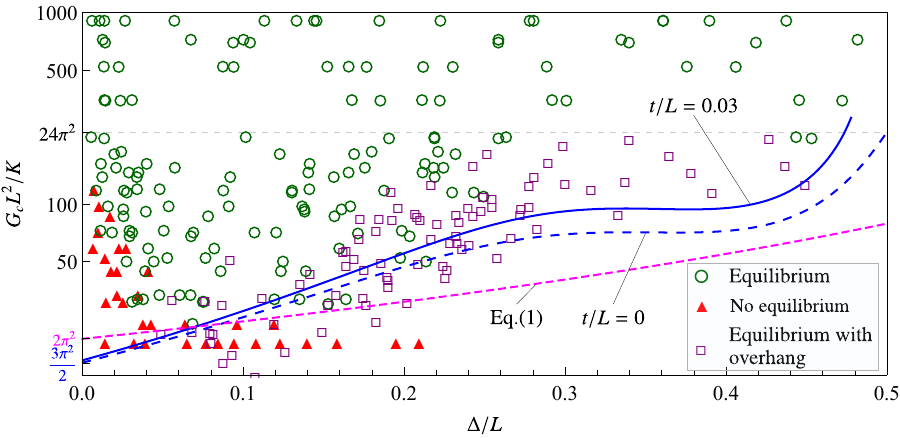}
        \caption{A comparison of experiments with the theory in the state space of an adhesive tape loop. The theoretical curves trace the points of limiting equilibria for a loop of a given normalized thickness $t/L$, and divide the state space such that the points above the curve admit equilibrium configurations, while the points below a curve do not. The scatter of points are experimental data, where the green circles and the red triangles represent loops that stay in equilibrium and the ones that unravel, respectively. The purple squares denote loops that opened up from the imposed state and settled in a different equilibrium with a smaller $\Delta$. The numerical value $t/L=0.03$ represents the  maximum value of length normalized thickness used in the experiments. The dashed blue curve represents the limiting equilibria of a tape loop with zero thickness. For such a curve,  $G_c L^2/K\rightarrow 3\pi^2/2$ as $\Delta\rightarrow 0$. The dashed magenta curve correspond to the scaling \eqref{Eqn:scaling1}.}
        \label{fig:GammaPlot1}
\end{figure}

Next we compare the experimental data in the state-space with the predictions from the scaling model and the Kirchhoff rod model, as shown in Fig.~\ref{fig:GammaPlot1}.
Both models predict curves of limiting equilibrium that divide the state space into two regions.
States that lie above the curves correspond to equilibrium, whereas the states which lie below correspond to no equilibrium.
The prediction from the scaling model (applicable to strips with zero thickness) is shown in a pink dashed line, whereas the two blue lines (dashed and continuous, corresponding to $t/L=0.0$ and $t/L = 0.03$) are computed from the Kirchhoff rod model.
The value of $t/L=0.03$ (the continuous line) is chosen from the highest value for non-dimensionalised thickness for which experimental samples were made.

For a strip of zero thickness, the two limits of overlap, i.e. $\Delta/L\rightarrow 0.0$ and $\Delta/L\rightarrow 0.5$, are of special interest.
In both these limits, the Kirchhoff rod model predicts the shape of regions $A$ and $B$ to be arcs of circles, with the radius of region $A$ being twice that of $B$ (since the modulus of the equivalent rod in region $A$ is twice the modulus of its constituents \eqref{eq:equivalent_moment_constitutive} when $t=0$).
The scaled curvatures for regions $A$ and $B$ for $\Delta/L\rightarrow 0$ can be computed as $\pi$ and $2\pi$, and consequently the scaled critical adhesive energy density can be obtained as the difference of their scaled bending energies $[(2\pi)^2 - \pi^2]/2 = 3\pi^2/2$.
Similarly, in the limit $\Delta/L \rightarrow 0.5$, the scaled curvatures are $4\pi$ for region $A$ and $8\pi$ for region $B$.
The scaled critical adhesive energy density can be computed as $[(8\pi)^2 - (4\pi)^2]/2 = 24\pi^2$. 

Data from experiments is plotted in Fig.~\ref{fig:GammaPlot1} on a log-linear scale for comparison with the theoretical predictions. 
Data points corresponding to equilibrium and no-equilibrium states appear to be delineated along lines similar to the theoretical prediction.
Equilibrium states with an overhang seem to appear closer to the curve of limiting equilibria.
While the agreement between experiments and theory appears good for moderate and large overlaps, there appears to be a notable discrepancy between the two for overlaps in the region $\Delta/L \lessapprox 4\%$ and adhesive strength $G_cL^2/K\lessapprox 115$.
We suspect that the discrepancy in this region between theory and experiments indicates greater sensitivity of the ATL to minor imperfections in the strength of the adhesive and experimental errors.
However, the theory and experiments appear in consonance for $\Delta/L \gtrapprox 4\%$ and $G_cL^2/K\gtrapprox 115$.

Loops with overhang were obtained when some loops opened up quasi-statically only to find equilibrium in a final new state.
\begin{figure}[t]
        \centering
        \includegraphics[width=0.98\textwidth]{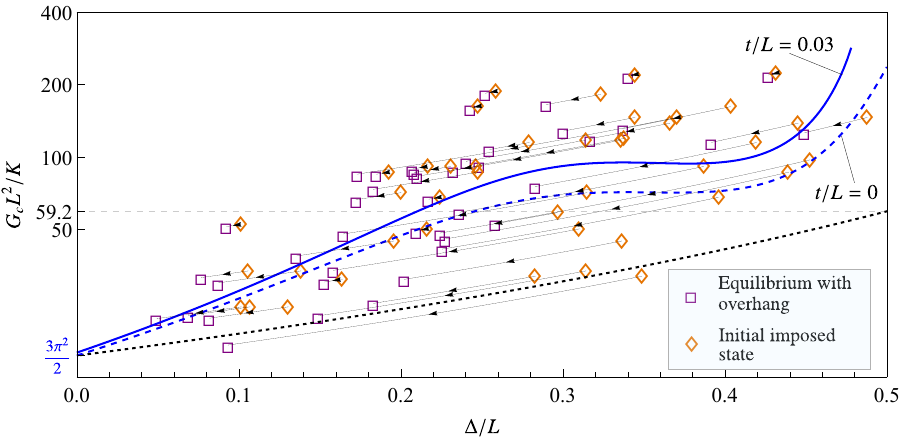}

        \caption{A subset of `Equilibrium with overhang' loops from Fig 4. is shown here. These loops were initially imposed with an overlap $\Delta_{initial}$ which are represented as orange diamonds and subsequently opened up and settled into equilibria represented by the purple squares. The black curves represent the theoretically predicted path going from one state to another, whose explicit parameterisation is given by $\{(\Delta-x)/(L-x), G_c(L-x)^2/K\}$. The dotted black line is a trace of this curve starting from a point chosen such that $G_c L^2/K\rightarrow 3\pi^2/2$ as $\Delta\rightarrow 0$.}
        \label{fig:GammaPlot2}
\end{figure}
The final new states are depicted in Fig.~\ref{fig:GammaPlot1} with square markers.
A natural question then arises: for which imposed states is it possible for a loop to peel and find a new equilibrium?
To answer this question, we compute the paths that an opening loop takes in the state space as the peeling progresses. 
Computing these paths is straightforward for a strip of zero thickness.
When such a strip opens up quasi-statically, the peeled off region is removed from the length of the loop, and the resulting curve followed in the state space 
can be written as,
\begin{align}\label{eq:quasi_static_path}
    \bc(\Delta_{excess}) = \left\{\frac{\Delta_0 - \Delta_{excess}}{L_0-\Delta_{excess}},\frac{G_c(L_0-\Delta_{excess})^2}{K}\right\} .
\end{align}
If the path described by the curve above crosses over from a no-equilibrium region to the equilibrium region of the state space for an initially imposed state $\{\Delta_0/L_0, G_cL_0^2/K\}$, we say that the loop is likely to settle into a configuration with an overhang.
If the curve $\bc$ does not intersect the curve of limiting equilibria, we conclude that the initially imposed state cannot find equilibrium by opening up, and it must unravel to its initially unbent state.

A comparison of \eqref{eq:quasi_static_path} with the experimental data is presented in Fig.~\ref{fig:GammaPlot2}.
Here, the initially imposed states are depicted by orange diamonds, and the final equilibrium that the loop settles into are shown with purple square markers.
The theoretically predicted paths for each of the initially imposed states are also shown.
It is quite remarkable that nearly all the final equilibria lie closely on their respective theoretically predicted quasi-static paths.
Since the paths are monotonically decreasing functions of $\Delta_{excess}$, it is possible to find curve below which no state can open up and settle into an equilibrium with overhang.
Such a curve is shown by the dotted black line in Fig~\ref{fig:GammaPlot2}.
All states below this curve must open up to their unbent state.
The experimental data shows that several states in the equilibrium part of the state diagram move to states with overhang lying in equilibrium region.
Several states cross over the band of limiting equilibrium bounded by the two blue curves and find equilibrium. A few experiments in the no-equilibrium part also settle into equilibrium. These outliers likely reflect the occasional imperfection in sample thickness, or perhaps occasional pinning of the contact line on imperfections. 

In the end, we note that the scaling relation \eqref{Eqn:scaling1} is actually a subset of the family of curves represented by \eqref{eq:quasi_static_path}.
Upon elimination of $\Delta_{excess}$ from the parametric form of the curve \eqref{eq:quasi_static_path}, one obtains the following relation
\begin{align}
    \frac{G_cL^2}{K}=y_0 \left(\frac{1}{1-\frac{\Delta}{L}}\right)^2\label{eq:temp}
\end{align}
where $y_0$ is the intercept of the curve on the $y-axis$.
The scaling relation is a curve \eqref{eq:quasi_static_path} such that it intercepts the y-axis at $2\pi^2$.

\section{Conclusion}\label{sec:conclusions}
We have experimentally and theoretically studied static equilibria of an adhesive tape loop formed by bending an initially straight strip to form a closed loop with some finite length close to its terminal ends overlapping.
Such a loop, depending on the strength of adhesion and the length of the overlap, may maintain equilibrium, unravel to go back to it unbent state, or open up quasi-statically to find equilibrium at an overlap different than the one imposed.
We performed experiments with strips made of PDMS, of varying lengths and thicknesses, and recorded the response of the loop after sufficient relaxation time. 
We derived, using simplifying assumptions on the geometry of the loop, a scaling relation between the length of the overlap and the adhesive strength required to hold the loop made from a strip of zero thickness.
Furthermore, we constructed a more detailed theoretical model for the adhesive tape loop using Kirchhoff rod theory.
The predictions made by the theory match well with the experimental data.
We also discuss the case where an adhesive tape loop might open up and settle into a different equilibrium, and in what cases would that be possible.
The theoretical model developed here could potentially be used to estimate the strength of self-adhesion in sticky materials simply by measuring the smallest overlap required to maintain an adhesive tape loop in equilibrium.

\begin{figure}[t]
        \centering
        \includegraphics[width=0.5\textwidth]{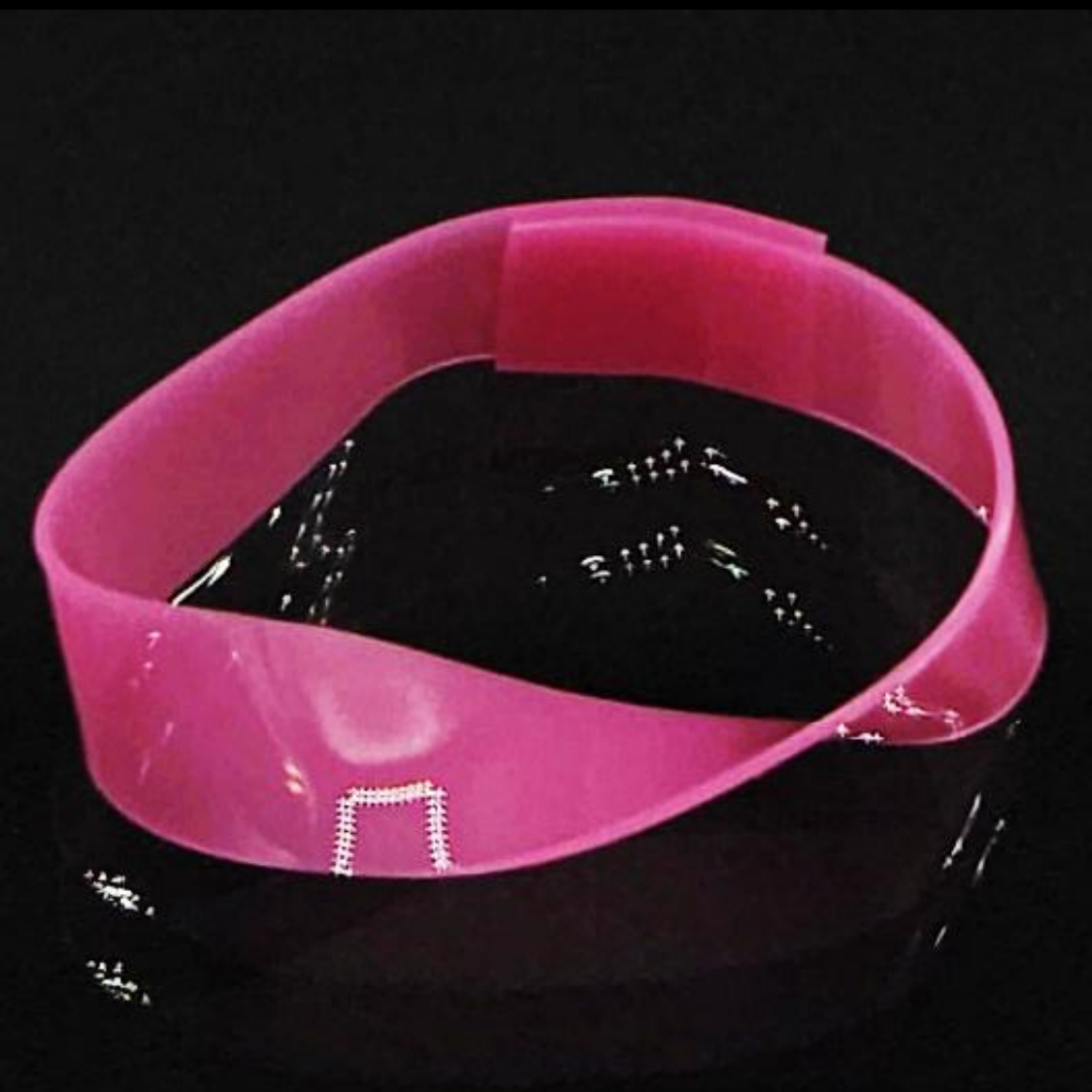}

        \caption{A loop containing a single $\pi$ twist.  A small overlap region in this case is sufficient to maintain a stable m\"{o}bius loop.  As the aspect ratio of the strip becomes smaller, the adhesive region becomes unstable and the loop opens.}
        \label{fig:mobius}
\end{figure}

An interesting extension of this problem could be to answer the same question in a situation where the strip is twisted by $\pi$ radians before forming the loop (see Fig.~\ref{fig:mobius}).
Such a twisted loop would form a Mobius band, which is a shape that is of great interest to both mathematicians and mechanicians.
We anticipate this problem to be far more formidable than the current case. 

\section{acknowledgements}
ABC would like to thank Wathsala Jayawardana for help with some of the imaging of loops, and Amara Arshad for contributing to the experiments.  ABC would like to gratefully acknowledge support from the National Science Foundation (NSF) (award no.: CMMI-2011681). 
HS thanks Dominic Vella for helpful discussions. HS gratefully acknowledges financial support for this work by the Science Engineering and Research Board (SERB) through grant SRG/2023/000079.

\appendix


\section{Image Processing details}\label{app:image_processing}

The images of the adhesive tape loops from the experiments are processed using built-in functions from {\texttt{Mathematica 12.0.0.0}}. 
The function \texttt{EdgeDetect} is used to detect the inner and outer boundaries of the loop. The function returns a list of coordinates $\{x_1,x_2\}$ in pixel scale that locate the edges/boundary of the loop. 
The list of coordinates are subsequently segregated to obtain two lists containing the coordinates of the inner and outer boundaries of the loop. 
A list of coordinates of the centerline is then computed by computing the midpoint between a point on the inner loop and the closest corresponding point on the outer loop. 
The total length of the centerline is computed by first applying a moving average on the data set to reduce the noise in the data.
The total length is then approximated by adding Euclidean distances between consecutive points on the centerline. 
Finally, the coordinates of the centerline are normalized by the total length for comparison with the theory.

Let $\bx_i\equiv\{x_1,x_2\}$ be the $i^{\textrm{th}}$ element of the list (with $N$ elements) of coordinates of the centerline curve.
We define a unit vector $\bc_i=(\bx_{i+1} - \bx_i)/\|(\bx_{i+1} - \bx_i)\|$ corresponding to each element $\bx_i$ with $i\in\{1,2,...\,,N-1\}$, pointing from $\bx_i$ to $\bx_{i+1}$.
A discrete equivalent of the tangent vector $\bt_i$ associated with the point $\bx_i$ can then be written as,
\begin{align}
    \bt_i = \frac{\bc_{i+1} + \bc_{i}}{2}\, ,\qquad i\in\{1,2,...\,,N-1\}\, .
\end{align}
The discrete curvature $\kappa_i$ associated with each point $\bx_i$ is then computed as the norm of the difference of the discrete tangent vector across consecutive points, divided by the Euclidean distance between the two points.
\begin{align}
    \kappa_i = \left|\frac{\bc_{i+1}-\bc_i}{\|\bx_{i+1} - \bx_i\|}\right|\, ,\qquad 
\end{align}
The list of discrete curvatures obtained is then smoothened by performing a moving average.

\section*{References}
\bibliographystyle{unsrt}

\end{document}